\documentclass[12pt]{article}
\usepackage{amssymb}
\usepackage{amsmath}
\usepackage{hyperref}
\begin{document}
\title{\bf Hydrodynamics of a  Black Brane in Gauss-Bonnet Massive Gravity}
\author{Mehdi Sadeghi\thanks{Email: mehdi.sadeghi@modares.ac.ir}  \hspace{2mm} and
        Shahrokh Parvizi\thanks{Corresponding author: Email: parvizi@modares.ac.ir}\hspace{1mm}\\
		{\small {\em  Department of Physics, School of Sciences,}}\\
        {\small {\em Tarbiat Modares University, P.O.Box 14155-4838, Tehran, Iran}}\\
       }
\date{\today}
\maketitle

\abstract{A black brane solution to a Gauss-Bonnet massive gravity is introduced. In the context of AdS/CFT correspondence, the viscosity to entropy ratio is found by the Green-Kubo formula. The result indicates violation of the well-known KSS bound as expected in a higher derivative theory. Setting mass zero gives back the known viscosity to entropy ratio dependent on the Gauss-Bonnet coupling, while without Gauss-Bonnet term, a nonzero  mass parameter doesn't contribute to the ratio which saturates the bound of $\frac{1}{4\pi}$.\\


\noindent \textbf{Keywords:} Gauss-Bonnet gravity, Massive gravity, Shear viscosity, Green-Kubo formula

\section{Introduction} \label{intro}

\indent For decades, various modifications of the Einstein gravity like Lovelock gravity \cite{Ref1}, brane world cosmology \cite{Ref2}, scalar-tensor theories\cite{Ref3}, and the so-called $ F(R) $ gravity have been proposed to address important problems in cosmology such as the cosmological constant, dark energy and dark matter. A common ingredient of these modified theories is the massless graviton. So it would be interesting to consider addition of a massive graviton to these theories {\cite{Ref7,Ref8}}. To study this new class of gravities it is important to find various spacetimes in different setups. One of the interesting setups is inclusion of the massive gravity in a higher derivative theory such as the Gauss-Bonnet gravity and looking for a black hole solution. This is what we do in section 2. In the context of massive Gauss-Bonnet gravity with a cosmological constant, we find a class of solution which includes black holes with either spherical or hyperbolic horizons and particularly a black brane solution with a flat horizon.    

On the other hand, it is well known from the AdS/CFT duality  \cite{Ref9,Ref10,Ref11} 
that quantum gravity, string theory, on an $ AdS_{d+1} $ background is dual to  a $d$-dimensional CFT, which is a non-gravitating theory. The latter can be described by hydrodynamics as an effective theory of QFT at large distances and time-scales  \cite{Ref12}. The AdS/CFT duality leads to fluid/gravity correspondence in this limit \cite{Ref13,Ref14,Ref15}. So it is worth to study the hydrodynamics of the QFT by the corresponding gravity theory via the AdS/CFT duality. The suitable solution to investigate the hydrodynamics is a black brane. 

The hydrodynamics equations are laws of conservation of energy and
momentum \cite{Ref12,Ref16},
\begin{align}
\nabla _{\mu } T^{\mu \nu } &=0, \\
T^{\mu \nu } &=(\rho +p)u^{\mu } u^{\nu } +pg^{\mu \nu }.
\end{align}

 According to dictionary of AdS/CFT, a black brane within the bulk is dual to a fluid on the boundary. It also implies that the Einstein equation in the bulk corresponds to the conservation of the energy-momentum equation on the boundary and the bulk metric is the dual field of the energy-momentum tensor of the boundary theory.

In the large wavelength limit where the hydrodynamics regime is valid, we expand the energy-momentum tensor as follows,
\begin{align}
& T^{\mu \nu } =(\rho +p)u^{\mu } u^{\nu } +pg^{\mu \nu } -\sigma ^{\mu \nu }\\
&\sigma ^{\mu \nu } = {P^{\mu \alpha } P^{\nu \beta } } [\eta(\nabla _{\alpha } u_{\beta } +\nabla _{\beta } u_{\alpha })+ (\zeta-\frac{2}{3}\eta) g_{\alpha \beta } \nabla .u]\\& P^{\mu \nu }=g^{\mu \nu}+u^{\mu}u^{\nu}, \nonumber
\end{align}
\indent where $\eta$, $\zeta $, $\sigma ^{\mu \nu }$ and $P^{\mu \nu }$ are shear viscosity, bulk viscosity, shear tensor and projection operator, respectively \cite{Ref15,Ref17,Ref18,Ref19,Ref20}. Here we are interested in the shear viscosity which can be derived by Green-Kubo formula \cite{Ref20,Ref21}. \\
\begin{equation}
\eta =\mathop{\lim }\limits_{\omega \to 0} \frac{1}{2\omega } \int dt\,  d\vec{x}\, e^{\imath\omega t} \left\langle [T_{y}^{x} (x),T_{y}^{x} (0)]\right\rangle =-\mathop{\lim }\limits_{\omega \, \to \, 0} \frac{1}{\omega } \Im G_{y\, \, y}^{x\, \, x} (\omega ,\vec{0}).
\end{equation}
Notice that the Green-Kubo prescription is independent of details of the bulk theory. Furthermore, the Gauss-Bonnet massive gravity has the advantage of being second order in derivatives of metric perturbation, so the Green-Kubo procedure preserves for the Gauss-Bonnet massive gravity.    \\
\indent In the following, after finding a black brane solution in the Gauss-Bonnet massive gravity, we calculate the shear viscosity to the entropy density ratio by applying Green-Kubo formula. The result shows that this ratio depends on both Gauss-Bonnet coupling and the graviton mass and violates KSS conjecture as expected in a higher derivative theory.
\section{Gravity Setup and the Black Brane Solution}
\label{sec2}

\indent The action of Gauss-Bonnet massive Gravity is as follows {\cite{Ref7,Ref8}},
\begin{align}\label{Action}
&I =\frac{1}{16\pi G_5}\int{d^5x\sqrt{-g}\Bigg[R+\frac{12}{L^2}+\frac{\lambda_{gb} }{2}L^2\mathcal{L}_{gb}+m^2\sum_{i=1}^4{c_{i}\mathcal{U}_i(g,f)}\Bigg]}\\
&\mathcal{L}_{gb}=R^2-4R_{\mu \nu}R^{\mu \nu}+R_{\mu \nu \rho \sigma }R^{\mu \nu \rho \sigma}\nonumber
\end{align}
where $ R $ is the scalar curvature, $f$ is a fixed rank-2 symmetric tensor and $m$ is mass parameter. In Eq.(\ref{Action}), $ c_i $'s are constants and $ \mathcal{U}_i $ are symmetric polynomials of the eigenvalues of the $ 5\times5 $ matrix $ \mathcal{K}^{\mu}_{\nu}=\sqrt{g^{\mu \alpha}f_{\alpha \nu}} $ given as {\cite{Ref7,Ref8}}
 \begin{align}\label{7} 
  & \mathcal{U}_1=[\mathcal{K}]\nonumber\\
  & \mathcal{U}_2=[\mathcal{K}]^2-[\mathcal{K}^2]\nonumber\\
  &\mathcal{U}_3=[\mathcal{K}]^3-3[\mathcal{K}][\mathcal{K}^2]+2[\mathcal{K}^3]\nonumber\\
  & \mathcal{U}_4=[\mathcal{K}]^4-6[\mathcal{K}^2][\mathcal{K}]^2+8[\mathcal{K}^3][\mathcal{K}]+3[\mathcal{K}^2]^2-6[\mathcal{K}^4]\nonumber
   \end{align}
The square root in $ \mathcal{K} $ means $ (\sqrt{A})^\mu_\nu(\sqrt{A})^\nu_\lambda=A^\mu_\nu $ and the rectangular brackets denote traces.\\    
We consider the following metric ansatz for a five-dimensional planar AdS black brane,
\begin{equation}
ds^{2} =-\frac{r^2N(r)^2}{L^2}f(r)dt^{2} +\frac{L^2dr^{2}}{r^2f(r)} +r^2h_{ij}dx^idx^j,
\end{equation}
A generalized version of $ f_{\mu \nu} $ was proposed in {\cite{Ref7,Ref8}} with the  form
$ f_{\mu \nu} = diag(0,0,c_0^2h_{ij})$, where $ h_{ij}=\frac{1}{L^2} $.

The values of $ \mathcal{U}_i $ are calculated as below,
\begin{align}\label{9} 
  & \mathcal{U}_1=\frac{3c_{0}}{r}, \,\,\,  \,\,\, \mathcal{U}_2=\frac{6c_0^2}{r^2},\,\,\,\,\mathcal{U}_3=\frac{6c_0^3}{r^3},\,\,\,\,\mathcal{U}_4=0\nonumber
   \end{align} 
Inserting this  ansatz into the action  Eq.(\ref{Action}) yields,
\begin{equation}\label{EOMI}
 I =\frac{1}{16\pi G_5}\int{d^5x\frac{3N(r)}{L^5}\Bigg[r^4\Bigg(1-f(r)+\lambda_{gb}f(r)^2\Bigg)+ m^2L^2c_0\Bigg(\frac{c_1 r^3}{3}+c_0c_1r^2+2c_0^2c_3r\Bigg)\Bigg]'}
\end{equation}
in which $'$ denotes derivative with respect to $r$. Equation of motion is given by variation of $N(r)$ {\cite{Ref22}},
\begin{equation}\nonumber
\Bigg[r^4\Bigg(1-f(r)+\lambda_{gb}f(r)^2\Bigg)+ m^2L^2c_0\Bigg(\frac{c_1 r^3}{3}+c_0c_1r^2+2c_0^2c_3r\Bigg)\Bigg]'=0
\end{equation}
$ f(r)$ is determined by solving the following equation,
\begin{equation}
r^4\Bigg(1-f(r)+\lambda_{gb}f(r)^2\Bigg)+ m^2L^2c_0\Bigg(\frac{c_1 r^3}{3}+c_0c_2r^2+2c_0^2c_3r\Bigg)=b^4
\end{equation}
in which $b$ is an integration constant. This yields two solutions,
\begin{equation}
 f_{\pm}(r)=\frac{1}{2\lambda_{gb}}\Bigg[1\pm\sqrt{1-4\lambda_{gb}(1-\frac{b^4}{r^4})-4m^2L^2\lambda_{gb}\Bigg(\frac{2c_0^3c_3}{r^3}+\frac{c_0^2c_2}{r^2}+\frac{c_0c_1}{3r}}\Bigg)\Bigg].
\end{equation}
Since black brane should have an event horizon, we choose $ f_-(r) $ of the solution above. 
Notice $ f_-(r) $ is zero at event horizon, $ r_+ $, so we have,
\begin{equation}
 b^4=r_+^4\bigg[1+m^2L^2\left(\frac{2c_0^3c_3}{r_+^3}+\frac{c_0^2c_2}{r_+^2}+\frac{c_1c_0}{3r_+}\right)\bigg]
\end{equation}

That's easy to show that $ N(r) $ is constant by variation of $ f(r) $ from Eq.(\ref{EOMI}). We use the dimensionless variable $ z=\frac{r}{r_+} $
\begin{equation}\label{metric}
 ds^{2} =-\frac{z^2r_+^2}{L^2}N^2f_-(z) dt^{2} +\frac{L^2 dz^2 }{z^2f_-(z)} +\frac{r_+^2z^2}{L^2} \sum_{i=1}^{3}dx_i^2
\end{equation}
with substituting $ b^4 $ in $ f_-(r) $,
\begin{eqnarray}
 f_-(z)&=&\frac{1}{2\lambda_{gb}}[1-\Gamma(z)]\\
 \Gamma(z)&=&\sqrt{1+\frac{4\lambda_{gb}}{z^4}\left[1-z^4+m^2L^2\Bigg(\frac{2c_0^3c_3}{r_+^3}(1-z)+\frac{c_0^2c_2}{r_+^2}(1-z^2)+\frac{c_1c_0}{3r_+}(1-z^3)\Bigg)\right]}\nonumber\\
\end{eqnarray}
In AdS/CFT correspondence, the speed of light in the boundary CFT is simply $c=1$. So that, for the black brane solution in the asymptotic region we have $\lim_{z \to \infty}{N^2 f_-(z)}=1$ to recover a causal boundary. By applying this criterion we will have,\\
\begin{equation}\label{causal}
N^2=\frac{1+\sqrt{1-4\lambda_{gb}}}{2}
\end{equation}
The temperature and the Hawking-Bekenstein entropy density follow,
\begin{eqnarray}\label{temp}
T&=&\frac{1}{2\pi}[\frac{1}{\sqrt{g_{rr}} } \frac{d}{dr} \sqrt{-g_{tt}}]|_{r=r_+ } =\frac{Nr_+^2}{4\pi L^{2}}f'_-(r_+)= \nonumber\\
\label{delta} &=&\frac{Nr_{+}}{\pi L^2}\Bigg[1+\frac{m^2L^2}{4}\Bigg(\frac{2c_0^3c_3}{r_+^3}+\frac{2c_0^2c_2}{r_+^2}+\frac{c_0c_1}{r_+}\Bigg)\Bigg] \equiv \frac{Nr_{+}}{\pi L^2}\Bigg[1+\Delta\Bigg]\\
\label{entropy} s&=&\frac{4\pi}{V}\int d^{3}x \sqrt{-g}=4\pi\left(\frac{r_+}{L}\right)^{3},
\end{eqnarray}
where in Eq. (\ref{delta}), the last equation defines parameter $\Delta$ and in Eq. (\ref{entropy}), we used $\frac{1}{16\pi G}=1  $ so $ \frac{1}{4\pi}=4G$.\\
We also generalize the metric ansatz to include spherical and hyperbolic, as well as planar horizons:
\begin{equation}
ds^{2} =-(k+\frac{r^2}{L^2}f(r))N(r)^2dt^{2} +\frac{dr^{2}}{k+\frac{r^2}{L^2}f(r)} +r^2h_{ij}dx^idx^j,
\end{equation}
where $ h_{ij}dx^idx^j $ is the following,
\begin{equation}
\begin{split}
k&=+1:\,\,\,\,\,d\Omega_3^2\,\,\,\,(\text{metric on} \,\,\,S^3),\\
k&=0\,\,\,\,\,:\,\,\,\,\,\frac{1}{L^2}\sum_{i=1}^{3}(dx_i)^2,\nonumber\\
k&=-1:\,\,\,\,\,d\Sigma_3^2\,\,\,\,(\text{metric on}\,\,\,H^3),\nonumber
\end{split}
\end{equation}
Note that for $k = \pm 1 $, the above line element has unit curvature. The horizon is determined by $ f(r_+)=-k\frac{L^2}{r_+^2}$.
\section{Shear Viscosity for the Black Brane}
 \label{sec4}
Let us start with the black brane solution
\begin{equation}\label{metricGB3}
ds^{2} =-\frac{r_+^2}{L^2}N^2\tilde{f} dt^{2} +\frac{L^2 dz^2 }{\tilde{f}} +\frac{r_+^2 z^2}{L^2} \sum_{i=1}^{3}dx_i^2,
\end{equation}
in which $ \tilde{f}=\frac{L^2}{r_+^2}f $. 

To find the shear viscosity, we perturb the background metric by $ h^{y}_{x}\equiv \frac{r_+^2 z^2}{L^2} \phi(t,\vec{x},z)$ 
\begin{equation}\label{metricGB}
ds^{2} =-\frac{r_+^2}{L^2}N^2\tilde{f} dt^{2} +\frac{L^2 dz^2 }{\tilde{f}} +\frac{r_+^2 z^2}{L^2}\Bigg( \sum_{i=1}^{3}dx_i^2+2\phi(t,\vec{x},z)dx_1dx_2\Bigg),
\end{equation}
Plugging in the action (\ref{Action}) and introduce Fourier modes of $\phi$, one finds 
 \begin{eqnarray}
 & & \phi(t,\vec{x},z)=\int \frac{d\omega dq}{(2\pi)^2}\phi(z,k)e^{-\imath\omega t+\imath qx}\; ,\hspace{1cm} k=(\omega,0,0,q) \nonumber \\
 \label{action2}
 & & I =\frac{-1}{2}\int{\frac{dzd\omega dq}{(2\pi)^2}\Bigg[K(\partial _{z}\phi)^2-K_2\phi^2+\partial _{z}(K_3\phi^2)}\Bigg],
 \end{eqnarray} 
 where $ \phi^2=\phi(z,k)\phi(z,-k) $ , $ \phi(z,-k)=\phi^*(z,k) $ and $K$'s functions will be introduced in the following. 
 We expand the metric up to second order of $ \phi $ and by using the general covariance symmetry\footnote{A general procedure is given in \cite{Cai:2008ph}.}, the action can be written as follows,
 \begin{equation}
  I =\frac{-1}{2}\int{\frac{dzd\omega dq}{(2\pi)^2}H\Bigg[g^{zz}(\partial _{z}\phi)^2+g^{tt}(\partial _{t}\phi)^2+m^2\frac{U(z)}{H}\phi^2+...}\Bigg],
  \end{equation}
  where,
 \begin{eqnarray}\label{54}
  K &=& Hg^{zz}=\frac{H\tilde{f}}{L^2}\nonumber\\
  \frac{K_2}{K} &=& \frac{g^{tt}}{g^{zz}}(-\omega^2-\frac{m^2U}{Hg^{tt}})
  =\frac{L^4}{r_+^2N^2\tilde{f}^2}(\omega^2-\frac{m^2 r_+^2N^2 U\tilde{f}}{L^2 H})\nonumber\\
  U &=& \frac{-Nc_0r_+^2}{4L^3}\left[3c_1r_+z^2+2c_0c_2z+108c_0^3c_4 z^{-1}\right]\nonumber\\
  K &=& z^2\tilde{f}(z-\lambda_{gb}\partial_z\tilde{f})\nonumber\\
  H &=& \frac{Nr_{+}^4}{L^3\Gamma(z)}\Bigg[(1-4\lambda_{gb})z^3
  -\lambda_{gb}L^2m^2\Bigg(\frac{2c_0^3c_3}{r_+^3}+\frac{2c_0^2c_2z}{r_+^2}+\frac{c_0c_1z^2}{r_+}\Bigg)\Bigg] 
   \end{eqnarray} 
in which we consider zero momentum $q=0$. Notice the surface term $K_3$ doesn't contribute to the viscosity. 
Mode equation can be derived from the action (\ref{action2}) as follows, 
\begin{equation} \label{main-eqGB}
K\phi^{''}+K^{'}\phi^{'}+K_2\phi=0
  \end{equation}
 Firstly, we solve the mode equation near the horizon. In near horizon limit we have,
  \begin{align}\label{K2K}
   &\frac{K_2}{K}\approx \frac{\omega^2}{B^2(z-1)^2}\\
   &\frac{K'}{K}\approx \frac{1}{z-1}
  \end{align} 
By comparing $K_2/K$ in (\ref{K2K}) and (\ref{54}) and ignoring the less singular term proportional to $m^2$, we obtain, 
\begin{equation}  
   B=\frac{N r_+}{L^2}\tilde{f}^{'}(1) 
\end{equation}
in which $'$ denotes derivative with respect to $z$. This equation is singular at the horizon $z=1$. The mode equation at near horizon is,
\begin{equation} 
\phi''(z) -\frac{1}{1-z}\phi'(z) +\frac{\omega^{2} }{B^2}\frac{1}{(1-z)^2}\phi(z)=0\, ,
\end{equation}
where its solution is $\phi(z)=(1-z)^{\beta }$, with
\begin{equation} \label{beta}
\beta =- \frac{\imath\omega}{B}
\end{equation}  
To solve the mode equation (\ref{main-eqGB}), we apply the following ansatz,
\begin{equation} \label{solutionGB}
\phi(z)={F}(z)^{\beta} (1-\beta h+O(\beta^{2} )),
 \end{equation} 
where $F(z)=N^2f(z)$. Inserting (\ref{solutionGB}) in (\ref{main-eqGB}) to first order of $\beta$, we obtain,
\begin{equation}
\Bigg(\frac{K{F}'}{{F}}\Bigg)'-Kh''-{K}' h'+m^2 U h=0 \,.
\end{equation}
It can be easily solved to find,
\begin{eqnarray} 
&(K\frac{F'}{F}- Kh')'+m^2 U h = 0 ,\\
& K\frac{F'}{F}- Kh'=A_1-m^2 \int^z U(z')h(z')dz' \\
& h=\log \frac{F}{A_2}-A_1\int^z\frac{1}{K}dz+m^2 \int^z\frac{dz_1}{K(z_1)}\int^{z_1}U(z_2)h(z_2)dz_2 
\end{eqnarray}
where $A_1$ and $A_2$ are integration constants. From the above equation, $h$ can be found perturbatively. However we need only near horizon and near boundary behavior of $h$ as we perform in the Appendix. For $ m^2=0 $ we have,
\begin{equation}\label{hzero}
h^{(0)}=\log \frac{F}{A_2}-A_1\int^z\frac{1}{K}dz
\end{equation} 
To be regular at the horizn it is enough to take $A_1=K'(1)$ and this preserves to all orders of $m^2$ as shown in the Appendix. 
We can read the retarded Green's function by expanding the action up to second order of source or the value of field on the boundary, 
\begin{equation}
S=-\frac{1}{2}\int \frac{d^{4}k}{(2\pi^4)}J_a(-k)\mathit{F}_a(k;z)J_a(k)\Bigg|_{z=\frac{1}{\epsilon}},
\end{equation}
then the retarded Green's function $ G_{R}(k) $ in momentum space for the boundary field dual is given by,
\begin{equation}
G_{R}(k)=-\mathop{\lim }\limits_{\epsilon \to 0}\mathcal{F}_{a}(k,z)|_{z=\frac{1}{\epsilon}},
\end{equation}
which can be calculated by the prescription given in {\cite{Ref23,Ref24}},
\begin{eqnarray} \label{Green0}
G_{R} (\omega ,\vec{0})&=& K \phi^{*}\partial _{z}\phi +K_{3}\phi^2=KF^{-\beta} (1+\beta h)\beta F^{\beta}(\frac{F'}{F}-h')\nonumber\\
 &=&\beta \Bigg(A_1-E\int^z U(z')h(z') dz' \Bigg)\Bigg|_{z\to \infty}=-\frac{\imath\omega}{B}A_1
\end{eqnarray}
In the first line $ K_{3}\phi^{2} $ is real and doesn't contribute to the viscosity, so we drop this term in the retarded Green's function. In the second line the integral term includes terms proportional to $c_1$, $c_2$ and $c_4$. A simple power counting indicates that the perturbative calculation of $h$ diverges for $z\rightarrow \infty$ unless $c_1=c_2=0$ and $c_4$ term converges to zero as we show in the Appendix. \\
Shear viscosity can now be found by using Green-Kubo formula as in the following,
\begin{equation} \label{etaGB}
\eta =-\mathop{\lim }\limits_{\omega \to 0} \frac{1}{\omega } \Im G_{R} (\omega ,\vec{0})= \frac{A_1}{B}=\frac{L^2}{r_{+}N\tilde{f}'(1)}K'(1)
\end{equation}
We can calculate $K'(1)$ as below,
\begin{equation}\label{Kprime1}
K'(1)=(Hg^{zz})'|_{z=1}= (H'g^{zz}+H(g^{zz})')|_{z=1}=H(g^{zz})'|_{z=1}=\frac{H\tilde{f}'}{L^2}|_{z=1}
\end{equation}
Then by substituting Eq.(\ref{Kprime1}) in Eq.(\ref{etaGB}), we will have, 
\begin{equation} \label{eq42}
\eta =\frac{A_1}{B}=\frac{H(z=1)}{Nr_{+}}
\end{equation}
and the ratio of shear viscosity to entropy density is,
\begin{equation} \label{eta-sGB}
\frac{\eta }{s} =\frac{1}{4\pi }\frac{L^3}{Nr_+^4} H(z=1).
\end{equation}
For the Einstein-Hilbert massive gravity $ H=\sqrt{-g}=\frac{Nr_+^4}{L^3} $, then the ratio of shear viscosity to entropy density is $ \frac{\eta}{s}=\frac{1}{4\pi} $ independent of the graviton mass. \\
We apply Eq.(\ref{eta-sGB}) for our black brane, with $H$ given in Eq. (\ref{54})
\begin{eqnarray} \label{eta-smassive}
\frac{\eta }{s} &=& \frac{1}{4\pi }[1-4(1+\Delta)\lambda_{gb}]  \\
 &=& \frac{1}{4\pi }[1-\frac{4\pi L^2 T \lambda_{gb}}{Nr_+}]
 \end{eqnarray}
where $ \Delta $  is in the following,
\begin{eqnarray}
\Delta &=& \frac{L^2m^2}{4}\Bigg(\frac{2c_0^3c_3}{r_+^3}+\frac{2c_0^2c_2}{r_+^2}+\frac{c_0c_1}{r_+}\Bigg)\\
&=& \frac{L^2m^2c_0^3c_3}{2 r_+^3}
\end{eqnarray}
where in the last line we put $c_1=c_2=0$. For $ m=0 $ we have, $ \frac{\eta}{s}=\frac{1}{4\pi }[1-4\lambda_{gb}] $ which is the well known result for the Gauss-Bonnet gravity.\\ 

\section{Conclusion}

\noindent In this paper we introduced a new solution to the Gauss-Bonnet massive gravity and showed that the lower bound $\eta/s=1/4\pi$ known as KSS conjecture \cite{Ref25}  violates for the black brane in this gravity. This is not surprising in the context of higher derivative theories {\cite{Ref26}-\cite{Roychowdhury:2014jqa}}. 

Results where exact in both the Gauss-Bonnet coupling and the graviton mass. It is interesting that for zero Gauss-Bonnet coupling the graviton mass disappears in the viscosity to entropy ratio and the result saturates the KSS bound, $1/4\pi$. On the other hand, inclusion of mass to the Guass-Bonnet looks like rescaling the Gauss-Bonnet coupling as  $\lambda_{gb} \rightarrow (1+\Delta)\lambda_{gb}$ in the viscosity to the entropy ratio with $\Delta$ a mass dependent parameter. Recalling the causal boundary criteria in (\ref{causal}), it indicates that $4\lambda_{gb}<1$. Now the rescaling of $\lambda_{gb}$ either decreases or increases the bound on $\eta/s$ ratio depending on the sign of $\Delta$. In any case, $1+\Delta$ is a positive parameter proportional to temperature as given in (\ref{temp}).


 \vspace{1cm}
 \noindent {\large {\bf Acknowledgment} }  Authors would like to thank Ali Imaanpur and Ahmad Moradpur for useful discussions.

 \vspace{1cm}
 \noindent {\large {\bf Note added} }  As this article was being completed, we received the preprint \cite{Hendi:2015pda} which has found the same solution as ours. 

\vspace{1cm}
\noindent {\large {\bf Appendix }} \label{App} \\

Here we study the behavior of $h$ function at the horizon and boundary limits. Firstly consider the following perturbative expansion of $h$ in powers of $m^2$:
\begin{equation}
h(z)=h^{(0)}(z)+m^2 h^{(1)}(z)+m^4 h^{(2)}(z)+\cdots 
\end{equation}
 where $h^{(0)}$ is given in (\ref{hzero}) and higher orders can be found by
\begin{equation}\label{perth}
h^{(n+1)}(z)= \int^z\frac{dz_1}{K(z_1)}\int^{z_1}U(z_2)h^{(n)}(z_2)dz_2 
\end{equation}
On the other hand, $F$ and $K$ are regular functions at $ z=1 $,
  \begin{eqnarray}
  F\approx F'(1)(z-1)\nonumber\\
  K\approx K'(1)(z-1) .
  \end{eqnarray}
  Thus to have non-singular $h^{(0)}$ at the horizon, $A_1$ is chosen to be $  A_1= K'(1)$. This result preserves at higher orders. Notice $U$ is a constant at the horizon, so taking $A_2=1$ then $h^{(0)}\approx C (z-1)+\cdots$ with $C$ some constant and
\begin{equation}
h^{(1)}(z)\approx \int^z\frac{dz_1}{K(z_1)}\int^{z_1}U(z_2)C (z_2-1)dz_2 =  \frac{U(1)C}{4K'(1)} (z-1)^2 
\end{equation}    
  then $h^{(n)}(z)=\frac{1}{(n!)^2}\left(\frac{U(1)C}{K'(1)}\right)^n (z-1)^{n+1}+\cdots$. This guarantees that $h$ vanishes smoothly near the horizon for any order of $m^2$.
  
  In the same way, we analyze the near boundary behavior of $h$. For $z \rightarrow \infty$, we have $K \sim z^3$, $F=N^2 f\sim 1$, and with $A_2=1$, one finds $h^{(0)}\sim 1/z^2$. Higher orders of $h$ can be derived perturbatively from Eq. (\ref{perth}). 
  
   Recall $U(z)$ includes terms proportional to $c_1 z^2$, $c_2 z$ and $c_4 z^{-1}$. It is easy to show that positive powers of $z$ in $U$ lead to increasing power of $z$ in each step of perturbation for $h$ and give a logarithmic divergence in few steps. So we have to take $c_1=c_2=0$. For $c_4\neq 0$ the integral converges at the boundary and indeed has zero contribution to the viscosity:
  \begin{eqnarray}
h^{(1)}(z)\sim \int^z\frac{dz_1}{K(z_1)}\int^{z_1}\frac{c_4 z_2^{-1}}{z_2^2}dz_2 \sim \frac{1}{z^4}
    \end{eqnarray}  
    Therefore $h^{(n)}(z)\sim 1/z^{2n+2}$ and the integral in Eq. (\ref{Green0}) becomes
 \begin{eqnarray}
\int^{z}U(z')h^{(n)}(z')dz' \sim \int^{z}\frac{(z')^{-1}}{(z')^{2n+2}}dz'\sim \frac{1}{z^{2n+2}}\rightarrow 0
 \end{eqnarray}  
    

\end{document}